\documentclass[prl,onecolumn,showpacs,superscriptaddress,nofootinbib]{revtex4}
\usepackage{graphicx}
\usepackage{color}
\usepackage{mathrsfs}

\begin{document}

\title{Some examples of quasiperiodic tilings obtained with a simple grid method}

\author{Jean-Fran\c cois SADOC}
\email{jean-francois.sadoc@universite-paris-saclay.fr}
\affiliation{Laboratoire de Physique des Solides (CNRS-UMR 8502), B{\^a}t. 510, Universit{\'e} Paris-Saclay, 91405 Orsay cedex, France}

\author{Marianne IMPEROR-CLERC}
\affiliation{Laboratoire de Physique des Solides (CNRS-UMR 8502), B{\^a}t. 510, Universit{\'e} Paris-Saclay, 91405 Orsay cedex, France}

\begin{abstract}
A grid method using tiling by fundamental domain of simple 2D lattices is presented. It refer to a previous work done by Stampfli in $1986$ using two tilings by regular hexagons, one rotate by $\pi/2$ relatively to the other. This allows to get a quasiperiodic structure with a twelve fold symmetry. The quasiperiodic structure is a tiling of the plane by regular triangles, squares and rhombuses. This can be extented to other examples of tilings by fundamental domain. Two other examples are proposed. The first example also based on the hexagonal lattice, but with grids defined by the fundamental rhombic domain formed by two regular triangles. The second example  presents the case of a square lattice with a square fundamental domain.
\end{abstract}

\maketitle


\section{Introduction }
Quasiperiodic tilings are more and more observed in experimental systems like 2D materials where they are directly linked to the superposition of periodic layers. A recent example is the case of dodecagonal graphene \cite{graphene}. In this context, the use of grid methods for building quasiperiodic tilings is highly relevant.

In a short paper published in 1986 Peter Stampfli \cite{stampfli} introduces the construction of a quasiperiodic tiling of the plane with regular triangles, squares and rhombuses having a global dodecagonal symmetry. He give a way to generate this tiling by a hierarchical decoration of tiles. But he introduces also a construction  derived from the overlap of two similar grids. The two grids are two periodic hexagonal tilings by identical regular hexagons. One grid is rotate by $\pi/2$ relatively to the other, so the two grids together have a dodecagonal symmetry.
Two years latter inspired by the Stampfli proposition, Korepin \cite{korepin} publish a more developed and general paper,  making a bridge between the grid method and the cut and project method.

The purpose of this paper is to give details of this construction by showing how it is possible to go from the two tiling by regular hexagons to the quasiperiodic dodecagonal tiling by squares, regular triangles and rhombuses.

It is interesting  to notice that this construction can be extended to other tilings. We present two examples. One uses a square lattice  to define a grid which is a tiling by squares and then a second grid obtained by $\pi/4$ rotation. The resulting quasiperiodic tiling is the Ammann-Beenker tiling by squares and rhombuses. The other example use the hexagonal lattice like in the Stampfli case, but grid are a tiling by rhombuses with $\pi/3,2\pi/3$ angles (two regular triangles). The quasiperiodic tiling contains squares, regular triangles rhombuses like in the Stampfli example, but also three-fold star-like additional tiles. It seems that the choice of different fundamental domains associated to lattices in order to get grids leads to a large choice of resulting quasiperiodic tilings.

\section{The two hexagonal grids of the Stampfli example}
The first grid is generated using an hexagonal lattice defined by the two base vectors:
$ \mathbf{e_1}=\{1,0\}, \mathbf{e_3}=\{-1/2,\sqrt{3}/2\}$. At each nodes of this lattice an hexagonal motif is reproduced leading to a tiling of the plane by hexagons. The vertices of this hexagonal motif are
$
 \{0 , \frac{\sqrt{3}}{3}\},
 \{-\frac{1}{2} , \frac{\sqrt{3}}{6}\},
 \{-\frac{1}{2} , -\frac{\sqrt{3}}{6}\},
 \{0 , -\frac{\sqrt{3}}{3}\},
 \{\frac{1}{2} , -\frac{\sqrt{3}}{6} \},
\{ \frac{1}{2} , \frac{\sqrt{3}}{6} \}
$.
Then the second grid is obtained rotating the first one by a $\pi/2$ rotation around the origin, so constructed relatively to a lattice having base vectors $(\mathbf{e_2},\mathbf{e_4})$ orthogonal to $(\mathbf{e_1},\mathbf{e_3})$. The choice of these notations for the four vectors
$(\mathbf{e_1}=\{1,0\},\mathbf{e_2}=\{\sqrt{3}/2,1/2\},\mathbf{e_3}=\{-1/2,\sqrt{3}/2\},\mathbf{e_4}=\{0,1\})$ refer to a previous publication  \cite{imperor} and is coherent with the notations used for the cut and project method (4D to 2D).

\begin{figure}[tbp]
\resizebox{0.8\textwidth}{!}{%
\includegraphics{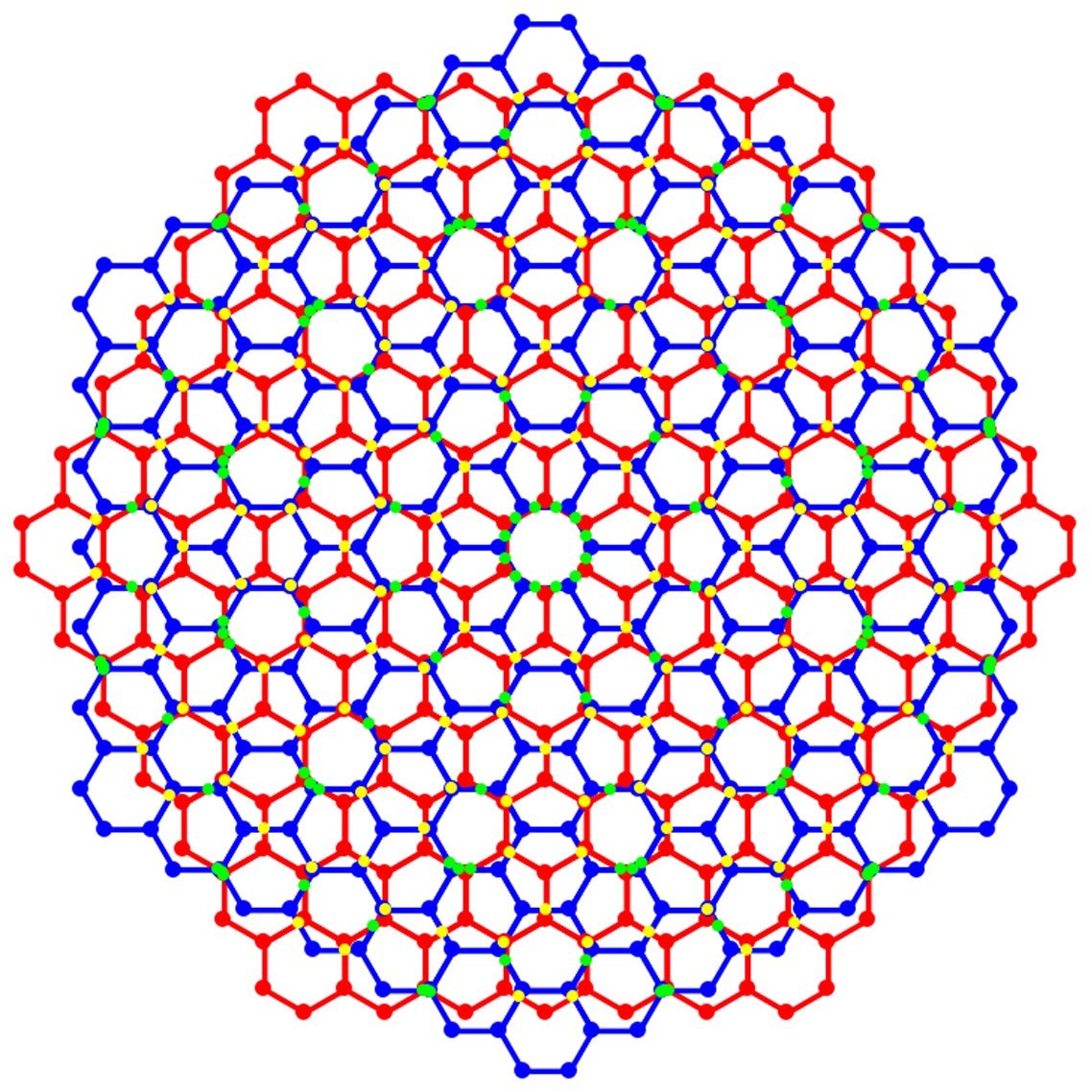}
}
\caption{The two grids, one represented in blue, the other in red. Vertices  between edges of hexagons are represented by points: red or blue for hexagons edges at hexagon vertices, yellow for two orthogonal edges, green for edges at $\pi/6$. Each
edge intersections will be associated with tiles: triangles for red or blue points, squares for yellow points and rhombuses for green points.}
\label{f1}
\end{figure}

Figure~\ref{f1} displays the two  grids of hexagons and also how edges of hexagons cut each others. They can cut at $\pi/3$ when the two edges belong to the same hexagon; they can be orthogonal, so with an edge belonging to one grid and the other to the other grid; they also can cut with a $\pi/6$ angle also from different grids. Stampfli states that each edge overlap define a tile of the quasiperiodic tiling: regular triangle if they are at $\pi/3$ angle, square for $\pi/2$ angle and rhombus for $\pi/6$ angle. Nevertheless Stampfli does not give a   clear demonstration. The rule which will give the tiling show that any overlap point corresponds to a tile even if the position of  points is not simply related to the tile position (it will even appear that an overlap point lies outside the related tile).

%
%
%
%
\section{Domains defined by the two grids}

Overlap of hexagons, one of each grid, define polygonal domains entirely covering the $2D$-plane. A domain have edges which are parts of edges of hexagons and have vertices which are the intersection points on hexagon edges, sometime from the same hexagon and otherwise from two hexagons from the two grids.

The two grids of hexagons (figure~\ref{f1}) are built by reference to two hexagonal lattices one related to the other by a rotation of $\pi/2$.  A domain is the common polygonal surface common to one hexagon of the first grid and another
of the other grid if both are sufficiently close to intersect. Hexagons are characterized by the lattice vectors positioning their centers. Call $(i,k)$ a first one define in the base $(\mathbf{e_1,e_3})$ and $(j,l)$  another in the base $(\mathbf{e_2,e_4})$   defining the two hexagons intersecting to form the domain we are considering. We attribute to each domain a reference point, located at the vector $(i \mathbf{e_1}+j \mathbf{e_2}+k \mathbf{e_3}+l \mathbf{e_4})/2$  which is the middle point between the two centers of the overlapping hexagons.

Notice that vectors $(i,k)$ and $(j,l)$ could be seen as vectors in a $E_{//}$ space projected from a space with higher dimension as it is often used in quasicrystal construction\cite{imperor}. Here is the relation between the grid method and the construction of quasicrystal by projection of higher dimensional space.

 The aim of this paper is to show that reference points of domains  are vertices of the quasi-periodic tiling.
%
%
\begin{figure}[tbp]

\resizebox{0.9\textwidth}{!}{%
\includegraphics{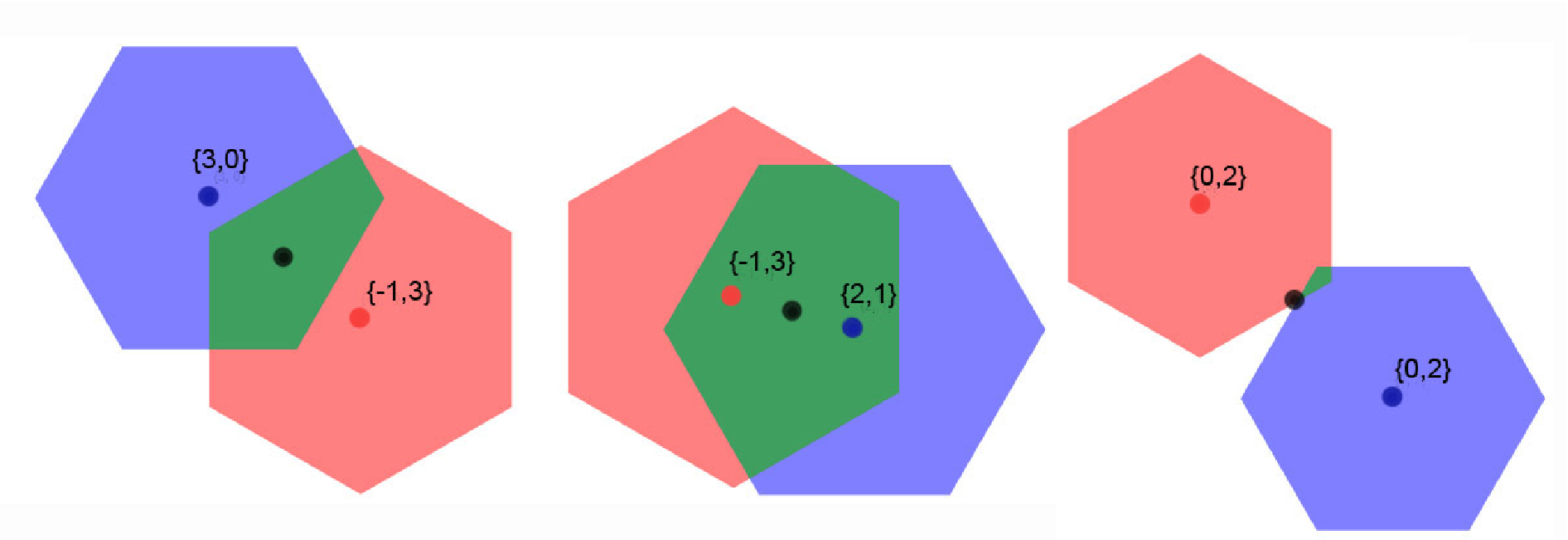}
}
\caption{Example of three different overlaps of two hexagons centered on nodes of the two lattices, one with coordinates $(i,k)$ on the base $(\mathbf{e_1,e_3})$ (red point) and one with coordinate $(j,l)$ on $(\mathbf{e_2,e_4})$ (blue points). Their common domain is shown with it charasteristic point (black point) corresponding to the middle point between the two centers of the overlapping hexagons. This point is a vertex of the quasiperiodic tiling.}
\label{f2}
\end{figure}
%
%

A polygonal domain is the overlap of two hexagons, one of the first grid (red $(i,k)$) and one of the other grid (blue $(j,l)$). We define, as reference point of this domain,  the point which bisect the vector joining the two centres of intersecting hexagons.

\begin{figure}[tbp]
\resizebox{0.5\textwidth}{!}{%
\includegraphics{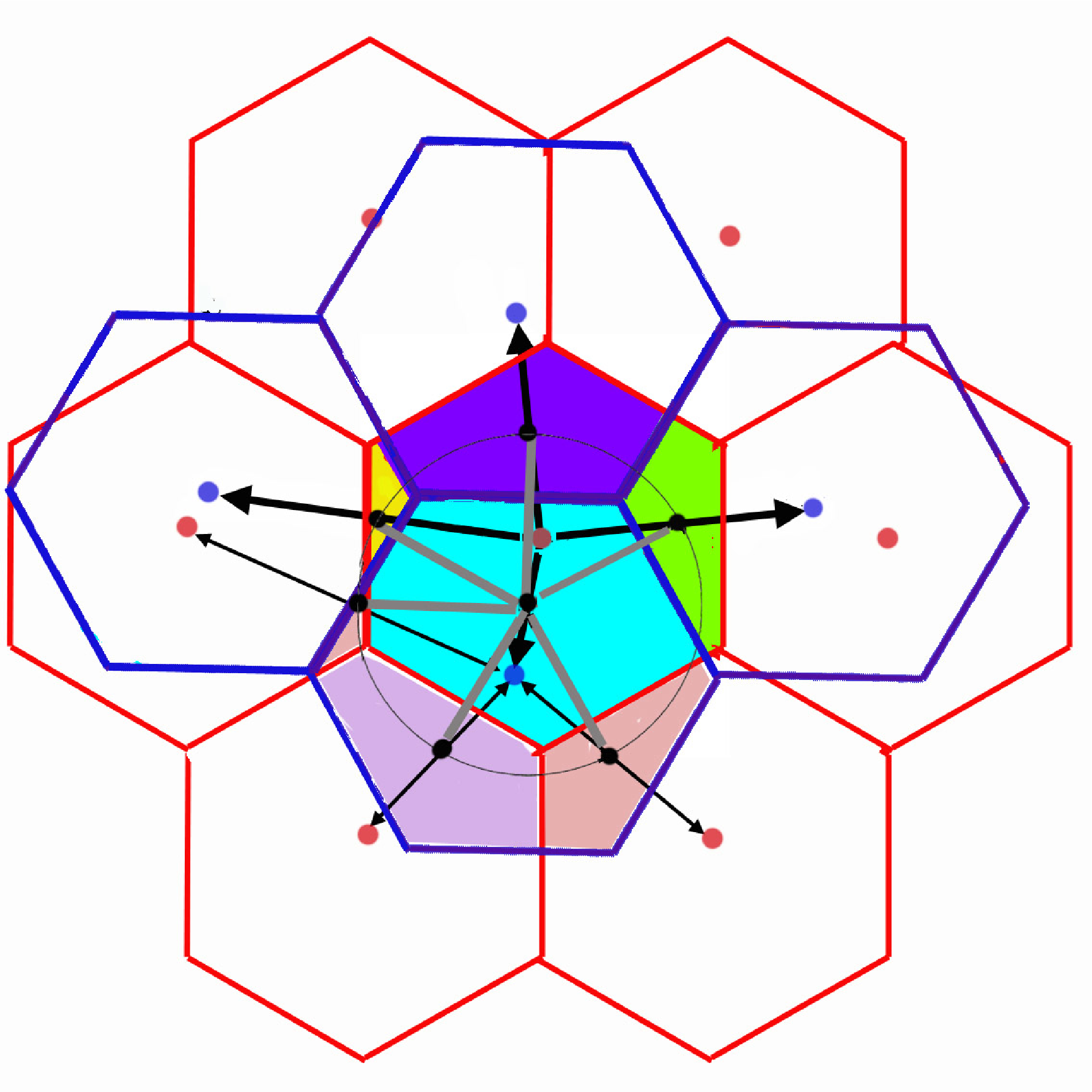}
}
\caption{Overlaping domains of two hexagons. The central red hexagon, centered on the central red point, intersect   blue hexagons, centered on vertices of the other lattice (blue points). Reference points (black points) of domains (vertices of the tiling) are mid points of vectors connecting the central red point to centres of blue hexagons.  Edges of the tiling (in grey) join black points, for instance the one between the blue and the green domains.
The vector joining the centre of the right and bottom blue hexagons, which are close neighbours, is a vector of the blue lattice of unit modulus.
So it appears that the length of the edge between the reference point of the blue domain and that of the green domain is half of this unit modulus.
That is the same for all edges of the quasiperiodic tiling.}

\label{f3}
\end{figure}

Consider a domain close neighbour to the previous one, so sharing one of it edges. Consequently this domain is necessarily the overlap of one of the previous hexagon (for instance the red one of the first grid $(i,k)$) and, in the other grid, a close neighbour with the hexagon   $(j,l)$; we can call it $(j',l')$. The vector joining $(j,l)$ and $(j',l')$ which are close neighbours in their own lattice, is a unit vector. So the length of the vector joining the two close reference points is the half of one unit vector defining lattices of grids (see the figure~\ref{f3}) .

The line joining the reference point of the first considered domain with this new one is now considered as an edge of the tiling. It length is half of the module of base lattice vectors used to construct the grids (said $1/2$). It is orthogonal to the common edge of hexagon $(j,l)$ and $(j',l')$. Extending this construction to all edges of the first domain we conclude that from the reference point of this domain, there are lines of length $1/2$ orthogonal to domain edges so possibly making $\pi/3$, $\pi/2$ or $\pi/6$ angles referring to angles possibly done by edges of domains (the red and blue points  or the yellow and green points of figure \ref{f1}).

%
%
%
%

 A reference point of a domain is connected by a  segment to a reference point of a neighbouring domain, so that the two domains are sharing a common edge. The important property is that the segment orthogonal to the shared edge has a length half of that of base lattice vectors, so all such segment have the same length.
 The angles appearing between segments joining close domains are the angles characteristic of regular triangles, squares or rhombuses. This is confirming that reference points of domains are vertices of a tiling by polygons whose edges have all the same modulus and angles are in the set $\pi/3$, $\pi/2$ or $\pi/6$. All tiles are necessarily  regular triangles, squares or rhombuses forming a quasiperiodic tiling.

 The quasiperiodic tiling can be constructed from the set of all reference points of all domains (which can be obtained using recent version (at least 13.1) of Mathematica \cite{mathematica} then we make a Delaunay triangulation of all these points. In this triangulation there are edges with length different from $1/2$ like square diagonals. By keeping in the triangulation only edges of length $1/2$, on obtains  the quasiperiodic tiling by regular triangles, squares and rhombuses.
%
%
\begin{figure}[tbp]
\resizebox{0.8\textwidth}{!}{%
\includegraphics{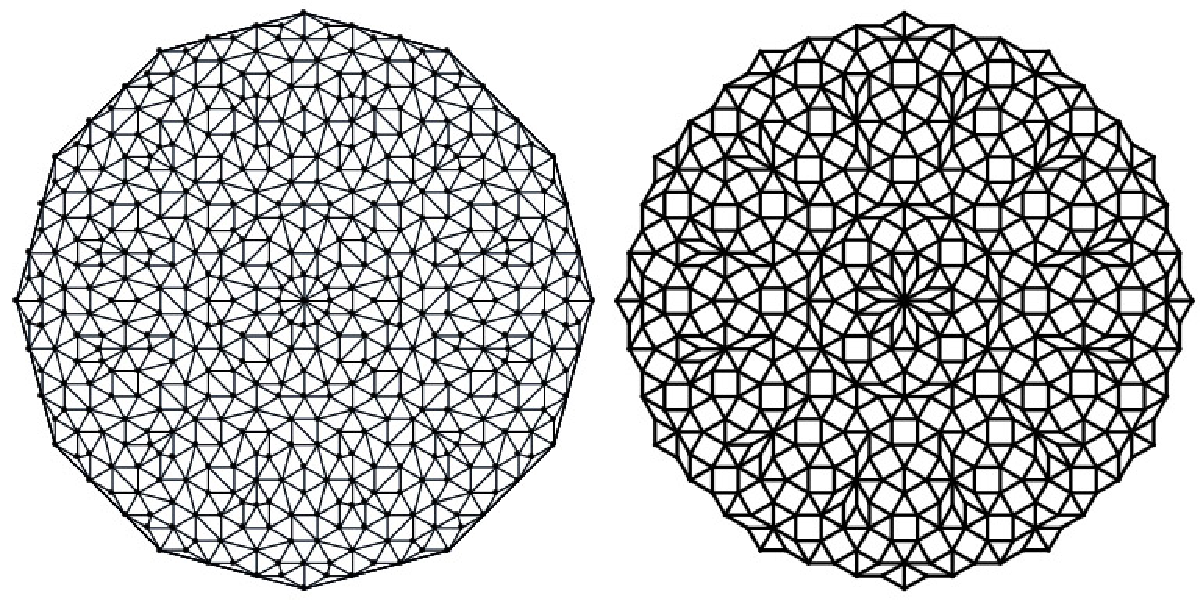}
}
\caption{Left; black points are charasteristic points of domains resulting from the overlap of hexagons. The figure is the Delaunay triangulation of this set of points (dual of Voronoi). Right; Tiles of the quasiperiodic tiling are regular triangles,  square  and rhombuses. This is derived from the Delaunay triangulation suppressing edges whose length is not $1/2$ (diagonal of squares and rhombuses and lines at the border). }
\label{f4}
\end{figure}
%
%
\section{An other example of this grid method: the octagonal tiling}
In place of the two hexagonal grids we consider instead two grids of squares, one with squares centered on the vertices of a square lattice with unit vector of unit length, the other grid being simply rotated around the origin by a $\pi/4$ angle. Then we construct all the polygonal domains corresponding to the overlap of squares of the two grids. The reference point of such a domain is the mid-point of square intersecting to form the domain.

%
%
\begin{figure}[tbp]
\resizebox{0.9\textwidth}{!}{%
\includegraphics{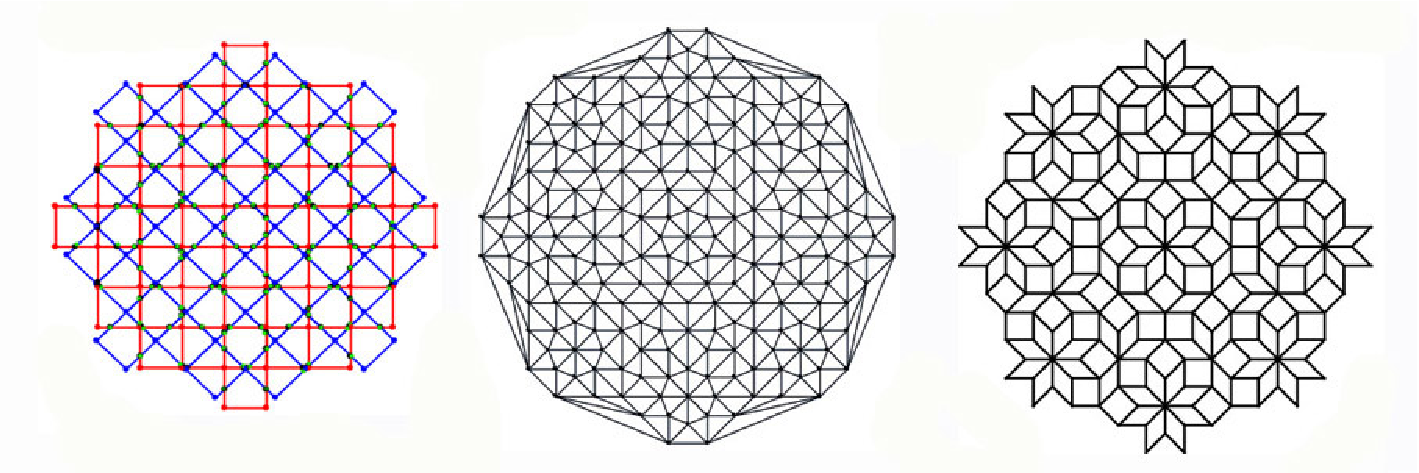}
}
\caption{Left; The two square grids with  intersections of edges. In the quasiperiodic tiling, vertices of grids lead to the two orientation of squares; overlaps between blue and red edges lead to rhombuses. Center; black points are charasteristic points of domains resulting from the overlap of squares. The figure is the Delaunay triangulation of this set of points (dual of Voronoi). Right; Tiles of the quasi-periodic tiling are  square  and rhombuses. This is derived from the Delaunay triangulation suppressing edges whose length is not $1/2$ (diagonal of squares and rhombuses and lines at the border). This is the Ammann-Beenker tiling. }
\label{f5}
\end{figure}
%
%

This set of points can be triangulated to form a Delaunay set. In this set some edges have a $1/2$ length (in term of base vector length); selecting these edges we get a tiling by squares and rhombuses which is quasiperiodic. In fact we recognize the well known Ammann-Beenker \cite{beenker,beenker2} tiling  obtained using inflation rules or by a different grid method.

%
%
\begin{figure}[tbp]
\resizebox{0.9\textwidth}{!}{%
\includegraphics{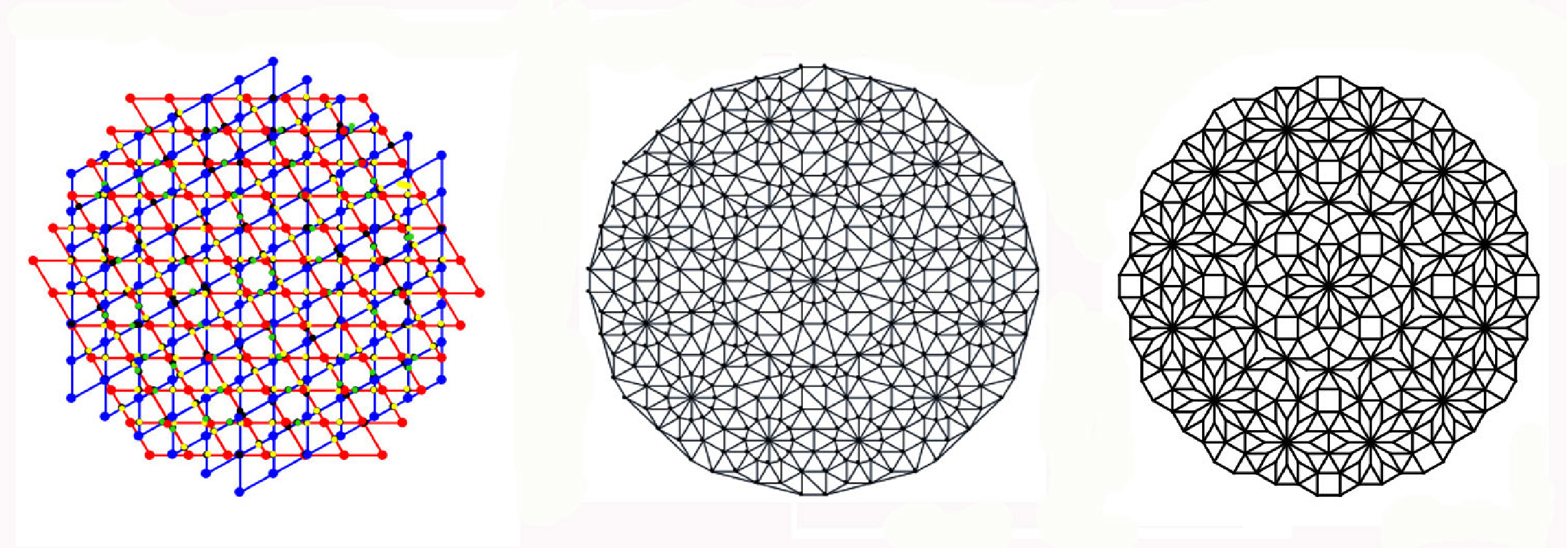}
}
\caption{Left: Grids which are rhombuses formed by gluing two regular triangles. Intersection of edges form $\pi/3$, $\pi/2$ or $\pi/6$ angles (with the same color code as in figure \ref{f1}), here again related to regular triangles, squares or rhombuses (or half rhombuses) in the quasiperiodic  tiling. Right: The quasiperiodic tiling resulting from this grid construction.  }
\label{f6}
\end{figure}
%
%

\section{An example with hexagonal lattices but with  less symmetric tiles for grids }
In order to get a grid the tile which is reproduced by the lattice has to be a fundamental domain of the lattice.
In the Stampfli example, the fundamental domain of the hexagonal lattice is a regular hexagon (a Voronoi cell of lattice nodes). But we can also choose a rhombus form by two glued regular triangles (the unit cell of the lattice). The resulting quasiperiodic tiling (figure \label{f6}) contain regular triangles, squares, rhombuses like the Stampfli one, but also star made of three half rhombuses radiating around a small triangle.

\section{Conclusion}
There are different methods which generate quasiperiodic tilings. The present one based on grids of regular tessellations of the plane is interesting as it makes a bridge between methods derived from projections of high dimension lattices and methods related to Moir\'e pattern.

The advantage of this method is that it is done directly in the plane and it consists in two simple steps. First the overlaping domains and the set of their reference points is build. Then the quasiperiodic tiling is obtained from the Delaunay triangulation of this set of points after removing extra edges. Both steps can be implemented using Mathematica software.

Using the grid method, the type of tiles in the quasiperiodic tiling depends on the intersections between edges of the tiles of the grid which have to be fundamental domain of the considered lattices. For instance, using hexagonal fundamental domain, intersections are with $\pi/6, \pi/2$ angles between grid tiles of the two grids, and with $\pi/3$ angles on grid vertices. The ratio between the different types of tiles is the ratio between the different types of intersection. An open question is that fundamental domain can have very complex shapes, eventually fractal\cite{damasco}, nevertheless the number of types of intersection remains probably small.

\end{document}